\documentclass[prb,twocolumn]{revtex4}

\usepackage{graphicx}
\usepackage{amsmath}
\usepackage{amssymb}
\usepackage{bm}

\usepackage{color}

\def\a{\alpha}

\def\e{\varepsilon}
\def\d{\delta}
\def\g{\gamma}
\def\l{\lambda}

\def\t{\tau}

\def\o{\omega}

\def\G{\Gamma}

\def\D{\Delta}

\def\ra{\rightarrow}

\def\pd{\partial}
\def\bk{{\bf k}}

\def\be{\begin{equation}}\def\ee{\end{equation}}
\def\bea{\begin{eqnarray}}\def\eea{\end{eqnarray}}
\def\nn{\nonumber}

\def\pref#1{(\ref{#1})}

\newcount\bozza \bozza=0
\ifnum\bozza=1
\newdimen\shift \shift=-2truecm
\def\lb#1{%
{\label{#1}\rlap{\kern\shift{$\scriptstyle#1$}}}}
\else\def\lb#1{\label{#1}} \fi

\begin{document}


\title{Robustness of the optical-conductivity sum rule in Bilayer Graphene}

\author{L.~Benfatto$^{1,2}$}

\author{S.G.~Sharapov$^{3,4}$}

\author{J.P.~Carbotte$^{4}$}

\affiliation
{$^1$Centro Studi e Ricerche ``Enrico Fermi'', via Panisperna 89/A, I-00184,
  Rome, Italy\\
$^2$CNR-SMC-INFM and Department of Physics, University of Rome ``La
  Sapienza'', P.le Aldo Moro 5, I-00185, Rome, Italy\\
$^3$ Department of Physics, Western Illinois University, Macomb, IL
61455, USA\\
$^4$Department of Physics and Astronomy, McMaster University,
        Hamilton, Ontario, Canada, L8S 4M1}

\date{\today }

\begin{abstract}
We calculate the optical sum associated with the in-plane conductivity
of a graphene bilayer. A bilayer asymmetry gap generated in a field-effect
device can split apart valence and conduction bands, which otherwise would
meet at two $\mathbf{K}$ points in the Brillouin zone. In this way one can
go from a compensated semimetal to a semiconductor with a tunable
gap. However, the sum rule turns out to be 'protected' against the opening
of this semiconducting gap, in contrast to the large variations observed in
other systems where the gap is induced by strong correlation effects.
\end{abstract}






\maketitle

\section{Introduction}
Sum rules on the conductivity have played an important role in the analysis
of optical conductivity data, which give information on electron
dynamics. In its simplest form, for an infinite free electron band, the sum
rule gives the plasma frequency which is independent of temperature,
interactions and impurity scattering.\cite{Benfatto:review,Carbotte:review}
In tight-binding models it is related to the second derivative with respect
to the momentum $k_{x,y}$ of the band dispersion
$\varepsilon_{\mathbf{k}}$, times the probability of occupation of the
state $|\mathbf{k}\rangle$.\cite{Carbotte:review,Benfatto:review} In the
special case where only nearest-neighbor hopping is present on a square
lattice, the sum-rule integral reduces to minus one half the kinetic energy
in appropriate units. As a consequence, in general it is affected by the
interactions present in the system, and these determine both its absolute
value and its variations with temperature.  This fact has been investigated
recently both experimentally and theoretically in the context of high-$T_c$
cuprate superconductors (see Ref.\
[\onlinecite{Benfatto:review,Carbotte:review}] for a review).  It follows
from the Sommerfeld expansion that the relative temperature variation of
the sum-rule integral in non-interacting tight-binding models is
proportional to $(T/t)^2$, where $t$ is the nearest-neighbor hopping. In
the experiments done in the normal state of cuprates, the temperature
variations of the the sum rule are much larger than this estimate, and in
some cases deviate from a simple $T^2$ law.  The discrepancy can be
attributed to correlation effects. For example, in the studies of the
Hubbard or $t-J$ model \cite{Toschi2005PRL,Haule2007EPL}, a new energy
scale emerges associated with the reduced width of the renormalized
band. This parameter replaces the hopping parameter $t$ in the $(T/t)^2$
dependence leading to its enhancement. It is shown in other theoretical
approaches that the sum rule integral measures instead a specific average
of the quasiparticle
lifetime\cite{Knigavko2004PRB,Marsiglio2006PRB,Benfatto2006PRB,Karakozov2006SSC}
normalized to the Fermi energy $\varepsilon_F$, and hence such optical
experiments ultimately probe correlation effects through lifetime
broadening.  These examples illustrate that optical data can provide
important insight into correlation effects.

Graphene, which is a single layer of graphite, has recently been isolated
and its properties investigated (see Ref.~\onlinecite{Geim:review} for a
review). Bilayer graphene\cite{Novoselov2006NatPhys} and thicker graphite
films are also now widely produced. Since graphene possesses truly
remarkable properties both from the technological and theoretical points of
view, there has already been considerable work done on monolayer and
bilayer graphene and also on related materials.  In particular, recent data
are available on the AC conductivity in the infrared region of
graphite\cite{Li2006PRB,Kuzmenko2007}, several layer epitaxial
graphite\cite{Sadowski2006PRL}, and on monolayer\cite{Jiang2007} and
bilayer\cite{Henriksen2008} graphene. There has also been much theoretical
work on the microwave and infrared conductivity of
graphene\cite{Gusynin2007AC,Falkovsky2007EPJB,Falkovsky2007PRB} (see
Ref.~\onlinecite{Gusynin2007review} for a review) and on a
bilayer\cite{Abergel2006PRB,McCann2007SSC,Nicol2008} and
multilayer.\cite{Nilsson2006PRL,Falkovsky2007PRB} Thus, it is of interest
to anticipate some general results on the optical-conductivity sum rule
behavior of these systems, which are likely to be tested experimentally in
the near future. The optical sum rules for the (in-plane) longitudinal and
Hall conductivities were studied for monolayer graphene in
Ref.[\onlinecite{Gusynin2007PRB}], where the unusual dependences of the sum
rules on temperature and chemical potential were revealed.  For example, at
the neutral Dirac point the linear dispersion law of quasiparticles leads
to a $T^3$ dependence of the sum-rule, instead of the $T^2$ law found in
tight-binding bands\cite{Benfatto:review,Carbotte:review}.

A specific and very useful feature of graphene, not available in ordinary
metals, is that it is possible to change the chemical potential $\mu$ and
thus the number of carriers by tuning the gate voltage $V_g$ ($V_g
\varpropto \mu^2$ for monolayer graphene) in a field-effect device. It was
shown\cite{Gusynin2007PRB} that in the limit $T=0$ the
concentration-dependent part of the sum rule goes like $(|\mu|/t)^3$.  For
a finite temperature $T \ll |\mu|$, there is a temperature dependent
correction to the sum of order $(|\mu|/t) (T/t)^2$ which is very much as in
the more ordinary $T^2$-law case discussed above.

In this paper we generalize the previous work to the case of bilayer
graphene.  In the case considered most often, a bilayer graphene
consists of two coupled hexagonal lattices.  The inequivalent sites
$\mathrm{A}1$, $\mathrm{B}1$ and $\mathrm{A}2$, $\mathrm{B}2$ on the
bottom and top graphene sheets are arranged according to Bernal
($\mathrm{A}2-\mathrm{B}1$) stacking: every $\mathrm{B}1$ site in
the bottom layer lies directly below an $\mathrm{A}2$ site in the
upper layer. The strongest inter-layer coupling between pairs of
$\mathrm{A}2-\mathrm{B}1$ orbitals, $\gamma_1$, changes the electron
dispersion from a linear to a quadratic form
\cite{McCann2007SSC,Guinea2007SSC}. In addition to this effect,
bilayer graphene is the only known material in which the electronic
band structure can be changed significantly simply by applying an
electric field perpendicular to the layers.\cite{Geim:review}
Indeed, while the unbiased bilayer system is a semi-metal, with
parabolic valence and conduction band touching each other at the
neutrality point, the system becomes a semiconductor when the two
layers are electrostatically inequivalent. The corresponding
semiconducting gap $\D$ between valence and conduction bands is
tunable continuously from zero to $\approx 0.2\,
\mbox{eV}$.\cite{Ohta2006Science,Castro2006,Geim:review} As shown
experimentally, the asymmetry gap $\Delta$ and the carrier
concentration can be controlled independently through an applied
gate voltage and chemical deposition of potassium or other atoms on
the vacuum side of the structure.\cite{Ohta2006Science,Castro2006}
Recently, similar results have also been obtained in a double-gated
structure made of bilayer graphene sandwiched in between two gate
electrodes \cite{Oostinga2007}. This effect has been understood by
means of both tight-binding
calculations\cite{McCann2006PRL,McCann2006PRB,Castro2006} and ab
initio calculations \cite{Min2007PRB}.

Beyond the clear potential impact that this effect has on device
applications, it is extremely interesting from the theoretical point
of view to understand how this field-induced band transition will
affect the conduction properties of the system. In light of the
previous discussion concerning the physics of cuprates, one could
ask if the opening of the gap due to the metal-insulator transition
in Hubbard-like models, or across the metal-superconductor
transition in superconducting models, share commonality with the
physics of bilayer graphene. As we shall demonstrate below, these
phenomena are drastically different. Indeed, we show that the sum
rule is essentially protected against the opening of the
semiconductor gap in this bilayer system. Despite the fact that $\D$
can be as big as a significant fraction of the Fermi energy, the
relative sum-rule changes from the unbiased to the biased system are
of order of $10^{-3}$. Thus, even though a general redistribution of
spectral weight is expected due to the opening of the gap, the total
sum rule is practically constant. One would have expected the large
band structure changes involved to lead to corresponding significant
changes in the optical sum. Here, however, we show that this is not
the case. The relative changes in optical spectral weight induced by
the opening of the gap must largely compensate each other, leaving
the total weight almost unchanged.

\section{Derivation of the sum rule}
For a generic electronic model described by the Hamiltonian $H$ the 
optical-conductivity sum rule is given
by:\cite{Benfatto:review,Carbotte:review}
\be \lb{common.rule} W(T)=\int_{-\infty}^{\infty} \mbox{Re}
\sigma_{xx}(\omega) d \omega =\frac{\pi e^2}{\mathcal{V}} \langle \tau_{xx}
\rangle, \ee
where $\mathcal{V}$ is the unit-cell volume and $\tau_{xx}$ is the
so-called mass tensor, which appears in the second-order expansion of
$H$ evaluated at finite vector potential ${\bf A}$:
\be \lb{j-tau.def} H(\mathbf{A})\approx H(0)  -
\sum_{\mathbf{n},\a=x,y} \left[ \frac{e}{c} A_\a(\mathbf{n})
j_\a(\mathbf{n}) - \frac{e^2}{2c^2} A_\a^2(\mathbf{n})
\tau_{\a\a}(\mathbf{n}) \right], \ee
and $j_\a$ is the current density in the $\a$ direction. The vector
potential $\mathbf{A}$ is introduced in the tight-binding Hamiltonian
(\ref{ham}) written below by means of the Peierls substitution,
$a_{2\mathbf{n}}^\dagger b_{2\mathbf{n}+\pmb{\delta}'}\rightarrow
a_{2\mathbf{n}}^\dagger\exp\left(- \frac{i e}{\hbar c}\int_{ \mathbf{n} +
\pmb{\delta}'}^{\mathbf{n}}{\bf A}d \mathbf{r}\right)b_{2 \mathbf{n}+
\pmb{\delta}'}$. Here $a_{2\mathbf{n}}$ and $b_{2\mathbf{n}+
\pmb{\delta}'}$ are the Fermi operators of electrons on $\mathrm{A}2$,
$\mathrm{B}2$ sublattices of the second layer. (Note that this Peierls
substitution corresponds to the 2nd layer, for the 1st layer it is
similar. The spin label is omitted.) The positions of $\mathrm{A}2$ and
$\mathrm{B}1$ atoms are denoted as $\mathbf{n}$ and they are connected to their
nearest neighbors on $\mathrm{B}2$ ($\mathrm{A}1$) sites by the three
vectors $\pmb{\d}'$ ($\pmb{\d}$). Accordingly, the general tight-binding
Hamiltonian for a bilayer reads
\bea \label{Hamilton-lattice}
H&=&-t\sum_{\mathbf{n},\pmb{\d}}(a^+_{1\mathbf{n}+\pmb{\d}}b_{1\mathbf{n}}+h.c.)
-t\sum_{\mathbf{n},\pmb{\d}'}
(a^+_{2\mathbf{n}}b_{2\mathbf{n}+\pmb{\d}'}+h.c.)\nn\\
\lb{ham}
&+&\g_1\sum_\mathbf{n}(a^+_{2\mathbf{n}}b_{1\mathbf{n}}+h.c.)+
\g_3\sum_{\mathbf{n},\pmb{\d}'}
(a^+_{1\mathbf{n}}b_{2\mathbf{n}+\pmb{\d}'}+h.c.),\nn\\
&-&\frac{1}{2}\D\sum_{\mathbf{n}}(a^+_{1 \mathbf{n}+\pmb{\d}}a_{1
\mathbf{n}+\pmb{\d}}+b^+_{1\mathbf{n}}b_{1\mathbf{n}}-(1\ra 2)),
\eea
where $t$ is the in-plane hopping parameter, and $\gamma_1$ the strongest
inter-layer coupling. When $\Delta=0$ this Hamiltonian is equivalent to the
Slonzewski-Weiss-McClure (SWM)
model\cite{Slonzewski1959PRev,Slonzewski1957PRev} for the bulk graphite,
provided that one retains only in-plane and $\gamma_1$ out-of-plane
hopping, and assumes no dispersion along the $c$ axis (perpendicular to the
planes)\cite{Mikitik2007}. As mentioned above, the connected sites
$\mathrm{A}2$ and $\mathrm{B}1$ lie directly below and above each other, so
that the magnetic field perpendicular to the bilayer does not affect this
coupling. This is not the case for the weaker $\mathrm{A}1-\mathrm{B}2$
coupling $\gamma_3$ included in the Hamiltonian (\ref{ham}) for
completeness. In what follows, however, we will neglect this term, whose
role is negligible as far as the longitudinal optical sum rule is
concerned.  Finally, the parameter $\Delta$ represents the asymmetry of the
on-site energies on the two layers, and it is induced by the electric field
applied perpendicular to the graphene bilayer. As we shall see below, it is
responsible of the splitting of the valence and conduction bands.

\begin{figure}[htb]
\includegraphics[scale=0.35]{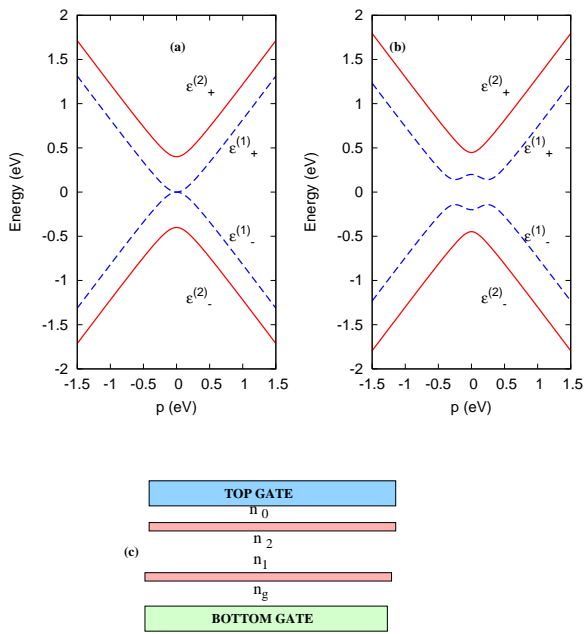}
\caption{(Color online) Top: bilayer band dispersion for $\D=0$ (a) and
  $\D=0.4$ eV (b) as a function of $p\equiv \hbar v_F k$. Bottom: general
  scheme of the experimental setup for a tunable-gap bilayer device
  (notation defined in the text).}
  \label{fig:1}
\end{figure}

The Hamiltonian \pref{ham} describes four
bands\cite{McCann2006PRB,Castro2006,McCann2007SSC,Guinea2007SSC}
$\e_{\pm}^{(\alpha)} (\mathbf{k})$, $\alpha = 1,2$
with
\bea
\e_\pm^{(\alpha)}(\mathbf{k})&=&\pm
\sqrt{\frac{\g_1^2}{2}+\frac{\D^2}{4}+|\phi(\mathbf{k})|^2
+ (-1)^\alpha \G}, \nn\\
\lb{eigenvalues}
\G&=&\sqrt{\frac{\g_1^4}{4}+|\phi(\mathbf{k})|^2(\g_1^2+\D^2)},
\eea
where the function $\phi(\mathbf{k}) = -t \sum_{\bm{\delta}_i} e^{i
\mathbf{k} \bm{\delta}_i} = -t \sum_{\bm{\delta}_i'} e^{-i \mathbf{k}
\bm{\delta}_i'}$. By expanding $\phi(\mathbf{k})$ around the two
inequivalent $\mathbf{K},\mathbf{K}'$ points in the Brillouin zone
one gets the usual linear dependence, $|\phi(\mathbf{k})|\approx
\hbar v_F |k|$, where the wave-vector $\mathbf{k}$ is measured from
$\mathbf{K}$ points and $v_F=\sqrt{3}ta/(2\hbar)$ is the Fermi
velocity, with $a$ the lattice constant ($\mathcal{V} = \sqrt{3}
a^2/2$). Using this form, one can easily see from
Eq.~\pref{eigenvalues} that at large momenta the four bands
reproduce the two Dirac cones $|\e_{\pm}^{(\alpha)}|=\hbar v_F k
\equiv |p|$ of each uncoupled layer. The low-energy band dispersion
is, however, drastically affected by the interplane hopping $\g_1$
and the asymmetry parameter $\D$. Indeed, at $\D=0$ for
$|\e_{\pm}^{(1)}| < \g_1/4$ the two low-energy bands are parabolic
with the dispersion $\e_{\pm}^{(1)}\approx \pm \hbar^2 k^2/2m$,
where the effective mass $m=\g_1/(2v_F^2)$, and touch each other as
in a semimetal (see Fig.~\ref{fig:1}a, where we used $t=3.1 \,
\mbox{eV}$, $\g_1=0.4\, \mbox{eV}$ \cite{McCann2006PRB,Castro2006}).
However, when $\D\neq 0$ the system becomes a semiconductor with a
gap $\tilde \D=\D\g_1/\sqrt{{\g_1^2+\D^2}} $ at the momentum
$p_m/\hbar v_F$, that corresponds to the energy $p_m^2=
(\D^2/{2})(\g_1^2+\D^2/2)/({\g_1^2+\D^2})$, see Fig.~\ref{fig:1}b.

Although the linear approximation for $\phi(\mathbf{k})$ is
sufficient for the analysis of the low-energy band structure and for
the numerical study of the mass tensor \pref{taufull} below, the
derivation of this tensor must be done using an unexpanded
expression for $\phi(\mathbf{k})$.\cite{Gusynin2007PRB} It follows
from the definition \pref{j-tau.def} that (the factor $2$ accounts
for the spin)
\be \lb{deftau} \t_{xx}=\frac{2}{\hbar^2 N}\sum_\mathbf{k} \left[
(a^+_{1\mathbf{k}}b_{1\mathbf{k}}+a^+_{2\mathbf{k}}b_{2\mathbf{k}})\frac{\pd^2
\phi^*(\mathbf{k})}{\pd k_x^2}+h.c.\right], \ee
where we kept the full momentum dependence of $\phi(\mathbf{k})$
($N$ is the number of unit cells). The thermal average $\langle
\t_{xx} \rangle$ is calculated from the the imaginary time Green's
function (GF) $G =- \langle T_\tau \Psi \Psi^{\dagger} \rangle $,
where we introduced
$\Psi^+=(a^+_{1\mathbf{k}},b^+_{2\mathbf{k}},a^+_{2\mathbf{k}},b^+_{1\mathbf{k}})$.
Then the averages $\langle a^+_{1\mathbf{k}}b_{1\mathbf{k}}\rangle$
and $\langle a^+_{2\mathbf{k}}b_{2\mathbf{k}}\rangle$ are
\bea \langle a^+_{1\mathbf{k}}b_{1\mathbf{k}}\rangle=T\sum_{i\o_n}
e^{-i\o_n 0^+}
{\hat G}_{41} (i \omega_n,\mathbf{k}), \nn\\
\lb{mats} \langle
a^+_{2\mathbf{k}}b_{2\mathbf{k}}\rangle=T\sum_{i\o_n}e^{-i\o_n 0^+}
{\hat G}_{23} (i \omega_n,\mathbf{k}), \eea where $i \omega_n$ is
the fermionic Matsubara frequency.
The corresponding elements ${\hat G}_{41}$ and ${\hat G}_{23}$ of the GF
${\hat G}$ can be found from the inverse GF\cite{McCann2007SSC}
\be
\lb{ginv}
\hat G^{-1}(i \omega_n,\mathbf{k})=\left(
\begin{array}{cccc}
z+\frac{1}{2}\Delta & 0 & 0 & -\phi^*(\mathbf{k})\\
0 & z-\frac{1}{2}\D& - \phi(\mathbf{k})& 0\\
0&  - \phi^*(\mathbf{k})& z-\frac{1}{2}\D& -\g_1 \\
- \phi(\mathbf{k})& 0& -\g_1 & z+\frac{1}{2}\D
\end{array}\right)
\ee
with $z = i \omega_n + \mu$. Then
\be {\hat G}_{41} (i \omega_n,\mathbf{k})=
\frac{\phi(\mathbf{k})[-|\phi(\mathbf{k})|^2+(\D/2-z)^2]}
{[z^2-(\varepsilon^{(1)})^2][z^2-(\varepsilon^{(2)})^2]}, \ee
and ${\hat G}_{23}={\hat G}_{41}(-\D)$.
Calculating the Matsubara sum in Eq.~\pref{mats} and using the
identity $\phi(\bk)(\pd^2 \phi^*(\bk)/\pd
k_\a^2)+c.c=-(a^2/3)|\phi(\bk)|^2$, we finally obtain
\be \lb{taufull}
\frac{\langle\t_{xx}\rangle}{\mathcal{V}}=-\frac{2a^2}{3\hbar^2
\mathcal{V} N} \sum_{\mathbf{k},\l=\pm, \a} |\phi(\mathbf{k})|^2
f(\e_{\l}^{(\a)}(\mathbf{k})-\mu)M(\e^{(\a)}_\l(\mathbf{k})),
\ee
where
\be M(\e_{\l}^{(\a)}(\mathbf{k}))= \left[\frac{1}{2}+ (-1)^\a
\frac{\g_1^2+\D^2}{4\G}\right]\frac{1}{\e_{\l}^{(\a)}(\mathbf{k})}
\ee
and $f(\e) = 1/[\exp(\e/T) +1]$ is the Fermi distribution. One can
verify that for $\g_1 = \D=0$ Eq.~(\ref{taufull}) reduces to the
doubled one-layer sum of Ref.~[\onlinecite{Gusynin2007PRB}].

The expression \pref{taufull} does not show any clear formal
resemblance to the kinetic-energy density of the bilayer system,
that one would define using the bands \pref{eigenvalues} as:
\be \lb{defkin} K=\frac{2}{\mathcal{V}N}\sum_{\mathbf{k},\l=\pm,\a}
\e_{\l}^{(\a)}(\mathbf{k}) f(\e^{(\a)}_\l(\mathbf{k})-\mu). \ee
Nonetheless, as we shall see these two quantities show approximately the same
doping dependence, satisfying again the general relation between sum
rule and kinetic energy, despite the complicated band evolution with
doping of the bilayer system.

\section{Sum rule as a function of doping}
As we already mentioned, it was shown
experimentally\cite{Castro2006,Ohta2006Science,Oostinga2007} that
one can tune independently the asymmetry gap $\D$ between the two
layers and the total carrier density (the value of $\mu$). This
corresponds to controlling the excess carrier density $n_{1,2}$
(difference between the densities of electrons and holes) in each
layer as schematically shown in Fig.~\ref{fig:1}c. The charge on the
surfaces below the bottom layer and above the top layer is $n_g$ and
$n_0$, respectively. In the experimental configuration of
Ref.~\onlinecite{Castro2006} $n_g$ is varied by changing the gate
voltage and $n_0$ by chemical deposition of dopants on top of the
upper layer (see also Ref.~\onlinecite{Ohta2006Science}), while in
the device of Ref.~\onlinecite{Oostinga2007} both $n_0$ and $n_g$
are varied, by using two independent gates. The resulting asymmetry
$\D$ between on-site energies in the two layers can be determined by
equating the voltage difference $\Delta/e$ between the plates of the
capacitor to its value evaluated from the charge density $e(n_0+ n_2
- n_g - n_1)$:\cite{McCann2006PRB,Castro2006,McCann2007SSC}
\be \lb{scr} \D=\frac{e^2d}{2\e_0}[n_0+ n_2(\D) - n_g - n_1(\D)
]=\frac{n_0ed}{\e_0} +\frac{n_2(\D)ed}{\e_0}. \ee
Here $d$ is the bilayer thickness, $\e_0$ is the permittivity of
free space and in the last equality we used that the total carrier
excess $n=n_1+ n_2=-(n_0+n_g)$. Since $n_1(\D=0) - n_2(\D=0)=0$, we
obtain that $\D=0$ at $n=-2n_0$. The gap closes when the system is
doped away from half-filling, as indicated by the ARPES measurements
of Ref.~\onlinecite{Ohta2006Science}. However, if the value of $n_0$
is tuned to be exactly at zero\cite{Oostinga2007}, the gap closes
exactly at the neutrality point, where the semimetal band structure
is reproduced. For finite $\Delta$ the excess carrier densities
$n_{1,2}$ are given by
\be
\begin{split}
n_1(\D,\mu)&=\frac{1}{N\mathcal{V}}\sum_{\bk,\a} \left.[
f(\e^{(\a)}_+ (\mathbf{k})-\mu)-f(\e^{(\a)}_+(\mathbf{k})+\mu) \right.\\
\lb{eqn1}
+& g^{(\a)}(\mathbf{k})\left.[f(\e^{(\a)}_-(\mathbf{k})-\mu)-f(\e^{(\a)}_+(\mathbf{k})-\mu)]\right],
\end{split}
\ee
where $g^{(\a)}(\mathbf{k})=
(\D/\e^{(\a)}_+(\mathbf{k}))(1/2+(-1)^\a|\phi(\mathbf{k})|^2/\G)$,
and $n_2(\D,\mu)=n_1(-\D,\mu)$. To determine selfconsistently the
dependence $\Delta$ on $n$, one has to solve simultaneously Eq.~\pref{scr}
and the equation $n_1(\D,\mu) + n_2(\D,\mu) =n$ for the chemical potential
$\mu$. Once $\D(n)$ and $\mu(n)$ are determined, one can compute $\langle
\tau_{xx}\rangle$ and $K$ as a function of $n$.

\begin{figure}[htb]
\includegraphics[scale=0.3,angle=-90]{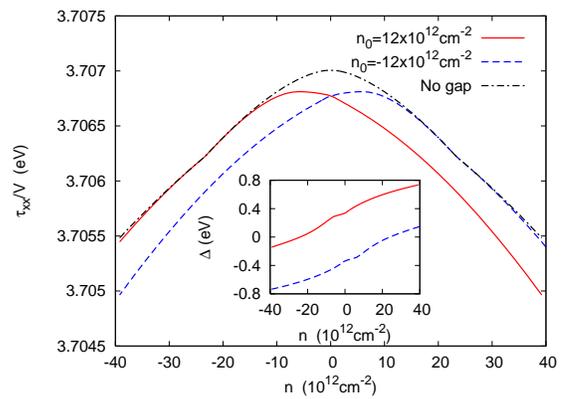}
\caption{(Color online) Doping dependence of the $T=0$ sum rule for a
  bilayer system with a doping-dependent gap $\D(n)$, shown in the
  inset. We show results for different values of $n_0$, given in units of
  10$^{12}$cm$^{-2}$. For comparison, we show the equivalent doping
  dependence of the sum rule when no gap is present between the two layers
  (dashed-dotted line).}
\label{sumt0}
\end{figure}

Following Ref.~\onlinecite{Ohta2006Science} we use $d=3.4$ \AA, and show
results for $n_0=0, n_0=\pm 12\times 10^{12}$ cm$^{-2}$. This choice is
suggested by the fact that in the measurements of Ref.
\cite{Ohta2006Science} the gap closes at $n=-2n_0$ given approximately by
$24\times 10^{12}$ cm$^{-2}$ (notice that $n$ expressed in units of
$10^{12}$ cm$^{-2}$ corresponds to $1.9\times 10^3 n_{uc}$, where $n_{uc}$
is the number of electrons per unit cell $n_{uc}$).  The doping dependence
of the $\langle \tau_{xx}\rangle$ at $T=0$ is shown in Fig.\ 2. As one can
see, even though the charging gap $\D$ changes consistently with doping
(see inset of Fig.~\ref{sumt0}), due to the screening effects of the
bilayer device, nonetheless the sum rule attains overall variations of
$\sim 0.1 \%$, which are undetectable from an experimental point of
view. In other words, despite the large optical-conductivity
spectral-weight redistribution associated with the opening of the gap
through the semimetal-semiconductor transition induced by doping, the sum
rule is not affected. It follows the same behavior that one would obtain in
(electrostatically) uncoupled layers where $\D=0$ at all doping (see the
dashed-dotted line in Fig.~\ref{sumt0}). In the regime where the gap is
relatively small (for example at negative doping for the $n_0=12\times
10^{12}$cm$^{-2}$ curve in Fig.~\ref{sumt0}) the sum rule follows
approximately the same doping dependence found in the single-layer case,
i.e. $[\tau_{xx}(n)-\tau_{xx}(0)]/{\cal V}\simeq -2a^2|\mu|^3/(9\pi
\hbar^2v_F^2)$. Since $2a^2/(9\pi \hbar^2v_F^2)\approx
10^{-2}\mbox{eV}^{-2}$, and $|\mu|^3$ has an overall variation of at most
$0.1 (\mbox{eV})^3$ in the considered doping range, one can easily get the
small sum-rule variation with doping. At larger gap values,
$\tau_{xx}(n)-\tau_{xx}(0)$ is slightly larger than expected in the
single-layer case, but is still too small to be detected experimentally. As
far as the temperature dependence is concerned we always found a $(T/t)^2$
variation, as expected due to the parabolic structure of the low-energy
bands.  Including other hoppings such as $\gamma_3$ in the Hamiltonian
(\ref{Hamilton-lattice}) can change the dispersion curve as discussed
in\cite{McCann2006PRL} (see also Ref.~\onlinecite{Mikitik2007}) for
energies less than $2\,\mbox{meV}$. While this can slightly affect the
overall value of the optical integral and of the self-consistent $\Delta$,
it does not change the main conclusion of our calculation, which focus on
the relative doping changes of the sum rule when a large semiconducting gap
$\Delta$ of order of a few hundred meV is opened. Having established that
the relatively large band-structure changes caused by the opening of the
gap $\Delta$ lead to a negligible changes in the optical sum, we expect
that the above mentioned hopping terms introduce negligible corrections as
well.

\begin{figure}[htb]
\includegraphics[scale=0.3,angle=-90]{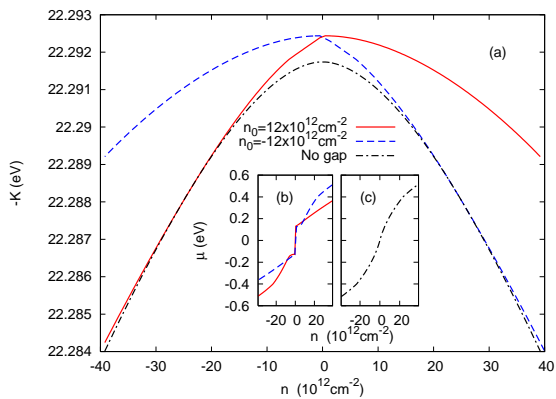}
\caption{(Color online) (a) Doping dependence of the $T=0$ kinetic energy
  for a bilayer system with a doping-dependent gap $\D(n)$ and without
  it. Observe the small variations of $K$ in the considered doping
  range. Insets: doping dependence of the chemical potential when a gap is
  present (b) or not (c), for parameters as in the main frame. }
\label{kinetic}
\end{figure}

Analogously, the kinetic energy, shown in Fig.~\ref{kinetic}, is not
much affected by the opening of the gap, and it attains a value
which is approximately 6 times the sum rule, $\tau/{\cal V}\sim
(1/6)|K|$. There are however two differences in the doping
dependence of the sum rule with respect to the kinetic energy.
First, we notice that the very small changes in the sum rule and
kinetic energy have the opposite trend: indeed, they both decrease
when the gap opens, in contrast to what is found for example at the
transition between a normal metal and a superconductor, where a
kinetic-energy increase corresponds to a sum-rule decrease and
vice-versa. This effect disappears when we set $\g_1 = \D=0$ in our
numerical work: indeed, in this case $\tau_{xx}$ reproduces exactly
twice the single-layer value computed in Ref.\
[\onlinecite{Gusynin2007PRB}], and the sum rule and the kinetic
energy track each other exactly  as the doping is varied. Second, it
is clear from Fig.\ \ref{kinetic} that the chemical-potential jump
at the neutrality point in the biased case (see inset (b) of Fig.\
\ref{kinetic}) is much more effective on the kinetic-energy doping
dependence. The two curves for $|K|$ at $n_0=\pm 12\times
10^{12}$cm$^{-2}$ in Fig.\ \ref{kinetic} show a kink at $n=0$ which
is clearly due to the $\mu(n)$ discontinuity at the neutrality
point. Indeed, in the case where no gap is present in the system
$\mu(n)$ goes smoothly across $n=0$, as does the kinetic energy.
Instead, the sum-rule curves in Fig.\ \ref{sumt0} are not much
affected by the $\mu$ discontinuity at $n=0$.

\section{Conclusions}

In summary we investigated the behavior of the optical sum in bilayer
graphene as a function of charge imbalance carrier density $n$ and
temperature. While a small temperature variation could be expected because
of the relatively small effect of temperature on the electronic
excitations, the negligible dependence of the sum rule on the carrier
density was not a priori expected. Indeed, despite the profound
band-structure changes induced by the change of doping, no significant
signature appears on the overall spectral weight. This implies that the
spectral weight lost in the gap must appear above it, and the two must
largely compensate for each other.  Observe that in order to be able to
test experimentally such a prediction one would need to define an
``experimental'' cut-off. Indeed, even though the relation
\pref{common.rule} is theoretically established by integrating the optical
conductivity to all frequencies, in practice an intrinsic cut-off is
provided by the frequency $\omega_c$ above which transitions to other
electronic bands not considered in the present study would
develop.\cite{Benfatto:review} According to optical studies on
graphite\cite{Taft1965PRev}, the optical sum rule saturates to the value
corresponding to one electron per atom (as due to the $\pi$ band
considered here) around 8 eV, while optical transitions coming from the
$\sigma$ bands appear around $15\, \mbox{eV}$. However, in the present case a
much smaller cut-off could be used, if one wants just to compare the
spectral-weight variations induced by the opening of the gap. Indeed, the
spectral-weight redistribution is expected to fall in a range of
frequencies of order $2\D$, as confirmed recently in Ref.\
\cite{Nicol2008}. Thus, an upper cut-off of order of $\omega_c\simeq 1-2$
eV should be enough to account for the gap-opening effects, and to test
experimentally the predicted robustness of the optical sum rule.

It is worth noting that in our calculation only the electrostatic
interaction between layers was included, since this is believed to
be the most important effect in the system. However, as in the case
of cuprates superconductors other mechanisms (electron repulsion or
electron-phonon interactions) could be at play and modify the sum
rule. Thus, the experimental verification of our prediction would
help understanding if other interactions need or need not be taken
into account in dealing with these systems.

\section*{Acknowledgments}
Useful discussions with V.P.~Gusynin, A.B.~Kuzmenko and E.J.~Nicol are
acknowledged.  J.P.C. and S.G.Sh. were supported by the Natural Science and
Engineering Research Council of Canada (NSERC) and by the Canadian
Institute for Advanced Research (CIFAR). L.B. acknowledges partial
financial support from Miur PRIN 2005, Prot. 2005022492.


\begin{thebibliography}{99}

\bibitem{Carbotte:review} J.P.~Carbotte and E.~Schachinger, J. Low Temp.
Phys. {\bf 144}, 61 (2006).

\bibitem{Benfatto:review} L.~Benfatto and S.~Sharapov, Fiz. Nizk. Temp. {\bf 32}, 700 (2006)
[Low Temp. Phys. {\bf 32}, 533 (2006).]





\bibitem{Toschi2005PRL} A.~Toschi,
M.~Capone, M.~Ortolani, P.~Calvani, S.~Lupi, C.~Castellani,
Phys. Rev. Lett. {\bf 95}, 097002 (2005). A.~Toschi and
M.~Capone, arXiv:0708.3485.

\bibitem{Haule2007EPL}
K.~Haule and G.~Kotliar, Europhys. Lett. {\bf 77}, 27007 (2007).

\bibitem{Knigavko2004PRB}
A. Knigavko, J. P. Carbotte, and F. Marsiglio, Phys. Rev. B. {\bf 70},
224501 (2004).

\bibitem{Marsiglio2006PRB} F. Marsiglio, Phys. Rev. B 73 064507 (2006);
Erratum to be published.

\bibitem{Benfatto2006PRB} L. Benfatto, F. Marsiglio, and J. P.
Carbotte, Phys. Rev. B {\bf 74}, 155115 (2006).

\bibitem{Karakozov2006SSC} A. E. Karakozov and E. G. Maksimov,
Solid State Comm. {\bf 139}, 80 (2006).


\bibitem{Geim:review} A.K.~Geim and K.S.~Novoselov, {\it Nature Materials}
{\bf 6}, 183 (2007).

\bibitem{Novoselov2006NatPhys} K.S.~Novoselov,
E.~McCann, S.V.~Morozov, V.I.~Fal'ko,
M.I.~Katsnelson, U.~Zeitler, D.~Jiang, F.~Schedin, and A.K.~Geim,
Nat. Phys. {\bf 22}, 177 (2006).

\bibitem{Li2006PRB} Z.Q.~Li,  S.-W.~Tsai, W.J.~Padilla, S.V.~Dordevic,
K.S.~Burch, Y.J.~Wang, and D.N. Basov,
Phys. Rev. B {\bf 74}, 195404 (2006).

\bibitem{Kuzmenko2007}
A.B.~Kuzmenko, E.~van~Heumen, F.~Carbone and D.~van~der~Marel,
arXiv:0712.0835v1 (unpublished).

\bibitem{Sadowski2006PRL} M.L.~Sadowski,
G.~Martinez, M.~Potemski, C.~Berger, and W.A.~de~Heer,
Phys. Rev. Lett. {\bf 97}, 266405 (2006);
Solid State Comm. {\bf 143}, 123 (2007).

\bibitem{Jiang2007} Z.~Jiang,
E.A.~Henriksen, L.C.~Tung, Y.-J.~Wang, M.E.~Schwartz, M.Y.~Han,
P.~Kim, and  H.L.~Stormer,
Phys. Rev. Lett. {\bf 98}, 197403 (2007).

\bibitem{Henriksen2008}
E.~A.~Henriksen, Z.~Jiang, L.~-C.~Tung, M.~E.~Schwartz, M.~Takita,
Y.~-J.~Wang, P.~Kim, H.~L.~Stormer, arXiv:0801.1861 (to appear on
Phys. Rev. Lett.). 


\bibitem{Gusynin2007AC} V.P.~Gusynin, S.G.~Sharapov and
J.P.~Carbotte, Phys.~Rev.~Lett. {\bf 96}, 256802 (2006), {\em ibid.}
{\bf 98}, 157402 (2007);
J. Phys.: Condens. Matter. {\bf 19}, 026222 (2007).

\bibitem{Falkovsky2007EPJB} L.A.~Falkovsky and A.A.~Varlamov,
Eur. Phys. J. B {\bf 56}, 281 (2007).

\bibitem{Falkovsky2007PRB} L.A.~Falkovsky and S.S.~Pershoguba,
Phys. Rev. B {\bf 76}, 153410 (2007).

\bibitem{Gusynin2007review} V.P.~Gusynin and S.G.~Sharapov,
Carbotte, Int. J. Mod. Phys. B {\bf 21}, 4611 (2007).

\bibitem{Abergel2006PRB} D.S.L.~Abergel, V.I.~Fal'ko,
Phys. Rev. B. {\bf 75}, 155430 (2007).

\bibitem{McCann2007SSC}
E.~McCann, D.S.L.~Abergel and V.~Fal'ko,  Solid State Comm. {\bf
143}, 110 (2007).

\bibitem{Nicol2008} E.J.~Nicol and J.P.~Carbotte, arXiv:0801.1836.

\bibitem{Nilsson2006PRL} J.~Nilsson,
A.H.~Castro~Neto, F.~Guinea, and N.M.R. Peres,
Phys. Rev. Lett. {\bf 97}, 266801 (2006).

\bibitem{Gusynin2007PRB} V.P.~Gusynin, S.G.~Sharapov and J.P.~Carbotte,
Phys.~Rev. B {\bf 75}, 165407 (2007).

\bibitem{Guinea2007SSC}  F.~Guinea, A.H.~Castro~Neto, and
N.M.R.~Peres, Solid State Comm. {\bf 143}, 116 (2007).

\bibitem{Castro2006}
E.V.~Castro, K.S.~Novoselov, S.V.~Morozov, N.M.R.~Peres,
J.M.B.~Lopes dos Santos, J.~Nilsson, F. Guinea, A.K. Geim, and A.~H.~Castro
Neto, \prl {\bf 99}, 216802 (2007).

\bibitem{Ohta2006Science}
T.~Ohta, A.~Bostwick, T.~Seyller, K.~Horn, and E.~Rotenberg,
Science {\bf 313}, 951 (2006).

\bibitem{Oostinga2007}
J.~B.~Oostinga, H.~B.~Heersche, X.L.~Alberto, F.~Morpurgo, and L.M.K.~Vandersypen,
Nature Materials {\bf 6}, 151 (2007).

\bibitem{McCann2006PRL}
E.~McCann and V.I.~Fal'ko, \prl {\bf 96}, 086805 (2006).

\bibitem{McCann2006PRB}
E.~McCann, \prb {\bf 74}, 161403(R) (2006).

\bibitem{Min2007PRB} H.~Min, B.R.~Sahu, S.K. Banerjee, and A.H.~MacDonald,
Phys. Rev. B {\bf 75} 155115 (2007).

\bibitem{Slonzewski1959PRev}
J.C.~Slonzewski and P.R.~Weiss, Phys. Rev. 109(2), 272 (1958).

\bibitem{Slonzewski1957PRev}
J.W.~McClure, Phys. Rev. 108, 612 (1957).

\bibitem{Mikitik2007} 
G.P.~Mikitik and Yu.V.Sharlai, arXiv:0709.1803.

\bibitem{Taft1965PRev} 
E.A.~Taft and H.R.~Philipp, Phys. Rev. {\bf 138}, A197 (1965).

\end{thebibliography}
\end{document}